\documentclass[journal=jacsat,manuscript=article]{achemso}

\usepackage{chemformula} 
\usepackage[T1]{fontenc} 

\author{Siyuan Wu$^{1,2}$}
\author{Ruijuan Xiao$^{1,2,3}$}%
 \email{rjxiao@iphy.ac.cn}
\author{Hong Li$^{1,2,3}$}
 \email{hli@iphy.ac.cn}
\author{Liquan Chen$^{1,2,3}$}
\affiliation{1. Institute of Physics, Chinese Academy of Sciences, Beijing 100190, People’s Republic of China\\2. School of Physical Sciences, University of Chinese Academy of Sciences, Beijing 100190, People’s Republic of China\\3. Yangtze River Delta Physics Research Center Co. Ltd, Liyang 213300, People’s Republic of China}
\title[An \textsf{achemso} demo]
  {LiY(SO$_4$)$_2$: A Superionic Material Synthesized by Superionic State Hidden in no-Superionic Literature}


\begin{document}


\begin{abstract}
A potential superionic material LiY(SO$_4$)$_2$ has been excavated from the published literatures because its synthesis method and experiment data implied it exists the superionic state. 
We use \textit{ab initio} calculation to analyzing the differences between solid state and superionic state. 
We found the diffusion of Li$^+$ from the lattice site to the interstitial site will change the nearest neighbor numbers of O atom from 4 to 8. 
In order to reduce energy, the reorientation of SO$_4^{2-}$ must exist accompany with the diffusion of Li$^+$ so the nearest neighbor number of O will keep about 5 in the superionic state. 
Our work not only presents an example for discovering materials from literatures based on prior knowledge but also reveals the micromechanism of cation-anion coupled dynamics for superionic state.
\end{abstract}

\section{\label{sec:Intro}Introduction}

\begin{figure}
\includegraphics[width=\textwidth]{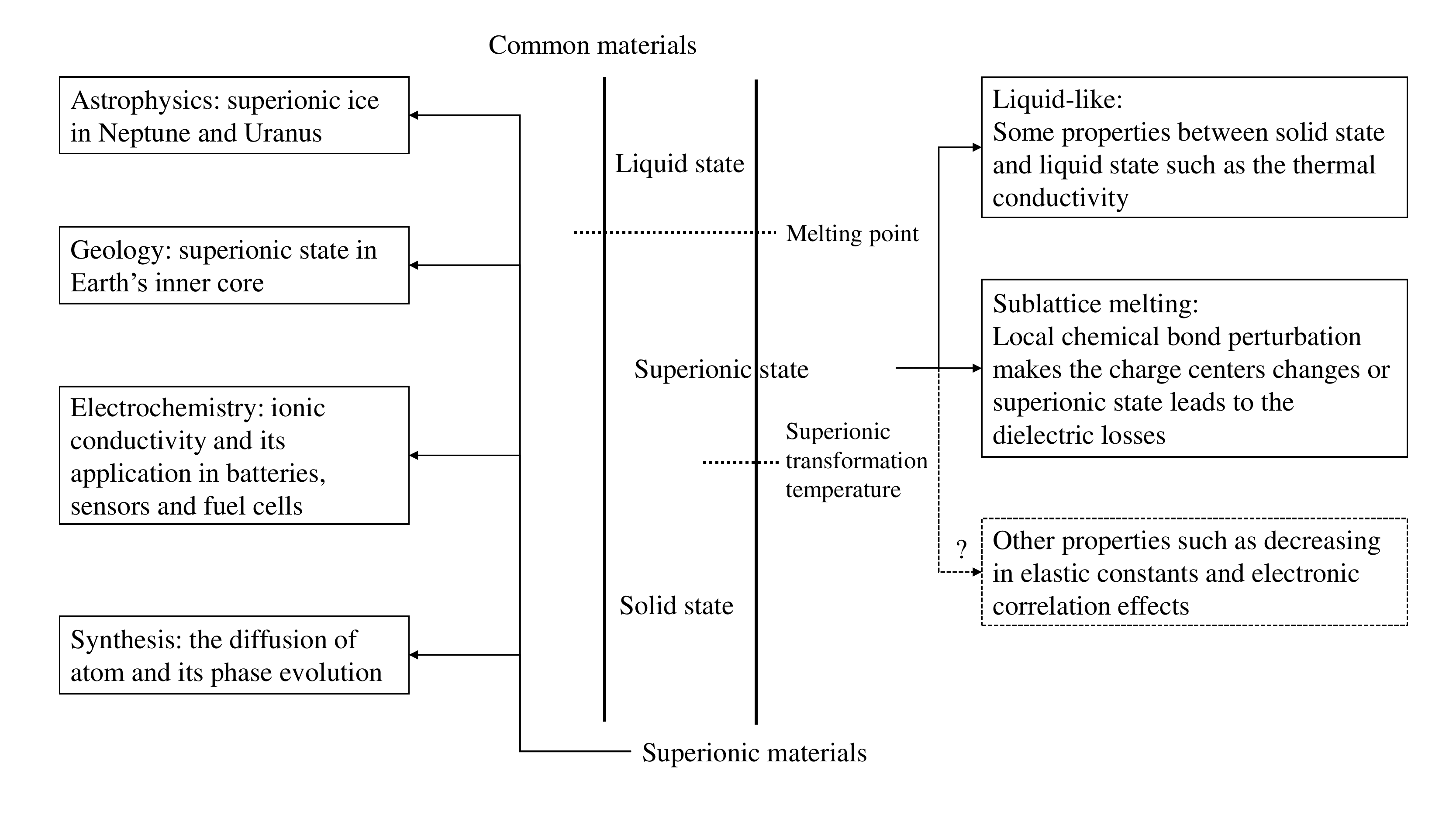}
\caption{\label{Fig_1}The concept of superionic state and its connections with other properties.}
\end{figure}

Superionic state or ionic diffusion \cite{Stephen_2004} has been widely researched in solid-state lithium batteries \cite{Tarascon_Nature_2001}, fuel cells \cite{Nguyen_1994_SOFC}, sensors \cite{ET_1995_sensors}, crystal growth and segregation \cite{CaoZX_JPCM_2011} and even in astrophysics \cite{Science_1999, SunJian_NP_2019} and geology \cite{HeY_Nature_2022}. 
It is interesting that these domains have different points although their research are all the superionic state as shown in Figure \ref{Fig_1}. 
They have different statements and concerns. 
Astrophysics pays attention to the superionic phase such as superionic ice \cite{Weck_PRL_2022,Alexander_NP_2021,Datchi_PRL_2020} especially when they found the abnormal magnetic field in Neptune and Uranus \cite{Science_1999}. 
They focus more on phase transformation temperature and order parameters. 
Geology and alloy pay more attention on diffusion coefficient \cite{Bai_NM_2022,HeY_Nature_2022} and their points of focus are in the phase evolution. 
In addition, the proton diffusion in the deep Earth \cite{Water2} and the deep interiors of Uranus and Neptune \cite{Water1} has been proved by \textit{ab initio} calculations which may be linked to the origin of water in Earth and giant planets.
For batteries, ionic conductivity and diffusion barrier are the first concern \cite{Xiao_PRL_2017,Kanno_NM_2011}. 
Of course, they still have the consensus on the superionic state: sublattice/partial melting \cite{SunJian_NP_2019,Alexander_NP_2021} and liquid-like \cite{Bai_NM_2022,HeY_Nature_2022}. 
In other words, if an article has the keywords like 'liquid-like' or 'sublattice/partial melting', it is probably linked with the superionic state. 
Of course, if there exist these keywords, whether in the physics or in the geology or others, its superionic state has been reported. The only advantage is that this will assist classifying the articles.\par

Although superionic state has a few portions in condensed matter physics and materials science, the character of partial melting makes superionic state superior especially in thermoelectric materials \cite{Cu2-xSe,AgCrSe2,Ag2Se} and dielectric materials \cite{LiB3O5,Rb2Ti2O5} as shown in Figure \ref{Fig_1}. 
As an intermediate state between solid state and liquid state, the physical and chemical properties also probably lie between them. 
In general, the thermal conductivity of solid is larger than that in the liquid state. 
Because of 'liquid-like' in the superionic state, there may be existing ultralow thermal conductivity due to the superionic state such as Cu$_{2-x}$Se \cite{Cu2-xSe}, AgCrSe$_2$ \cite{AgCrSe2} and Ag$_2$Se \cite{Ag2Se}. 
Superionic state is originated from the low-energy mode \cite{LEM_in_SSI} and Muy et al. \cite{EES_2018} associated the fast ion conductors with low center of lithium phonon density of states so the relevance between superionic state and thermal conductivity should exist. 
In addition, even below the superionic transformation temperature, the larger bond fluctuation in superionic materials \cite{bond1,bond2} may induce the changes of charge center which is connected with dielectric property and above the superionic transformation temperature, superionic state may lead to the dielectric losses \cite{LiB3O5,LLTO}. 
For example, LiB$_3$O$_5$ \cite{LiB3O5} exhibits the dielectric losses due to the ionic conductivity. 
Interestingly, Li$_{0.34}$La$_{0.56}$TiO$_3$ was considered as a superionic conductor by Yoshiyuki and Chen due to its large dielectric loss and dielectric relaxation \cite{LLTO}. 
Some materials such as Rb$_2$Ti$_2$O$_5$ \cite{Rb2Ti2O5} own high dielectric constant due to ion migration and accumulation. 
Besides, the superionic state has been considered been connected with abnormal decrease in $c_{33}$ and $c_{44}$ elastic constants in K$_3$Nb$_3$O$_6$(BO$_3$)$_2$ \cite{KNOBO} and polaronic hopping in KTiOPO$_4$ \cite{KTOPO}. 
These researchers indicate that we can connect other properties with superionic state at least in thermoelectric materials \cite{Cu2-xSe,AgCrSe2,Ag2Se} and dielectric materials \cite{LiB3O5,Rb2Ti2O5} from the view of atom thermal perturbation. 
Apart from this, the electronic correlation effects are considered as the reason for superionic state in superconductor RBa$_2$Cu$_3$O$_{7-\delta}$ \cite{R_PRB_1997}. 
Due to the different temperature range, most superconducting literatures little refer to the superionic state except a few materials such as Li$_6$Al \cite{SunJian_PRL_2022} and AsLi$_7$ \cite{ZRQ_2022_PRB} but in Ref. \cite{R_PRB_1997} it seems some connections between them.
In a word, superionic state should not only been researched as an intermediate state but also been existed in more fields.\par

Above discussions present the relevance between superionic state and other properties and it will instruct us to seek new superionic materials. 
Although there are some unbiased methods such as high throughput calculations and experiments even automation \cite{JPCL_2011,Curtarolo_CMS_2010,Burger_Nature_2020} to seek new novel materials from materials database, it seems need a lot of resources because they don't refer to the reported physical property. 
If we can seek a superionic material through bias method based on above discussions, it is more efficient. 
In fact, it is highly possible because the superionic state is the property connected with structure and thermodynamic property. 
Most articles with new materials will present the structure parameters and thermodynamic property. 
Some keywords such as 'liquid-like', 'site disordered' and such properties especially in thermoelectric materials and dielectric materials may imply its superionic state. 
In addition, the solid state reaction is an example of utilizing the ionic diffusion and superionic state \cite{SSR,AM} and we can infer its superionic state based on it.\par

In this letter, we will present an example LiY(SO$_4$)$_2$ which is first reported as nonlinear optical materials recently in Ref. \cite{LYSO}. 
We found its synthetic method and experimental results implied its superionic state intensely. 
We use \textit{ab initio} molecular dynamics and data mining to analyze its superionic character and reveal it exists superionic state above 900K which is in accord with our inference from Ref. \cite{LYSO} and the reorientation of SO$_4^{2-}$ will coexist with the diffusion of Li$^+$ in order to reduce the changes of bonds.

\section{\label{sec:res}Results and Discussion}
\subsection{Potential Character in the Superionic State}
Superionic state is an intermediate state between solid state and liquid state and part of atoms act as solid while others act as liquid. 
In the Ref. \cite{LYSO} the authors synthesized it at 873K for 12h with Li$_2$SO$_4$$\cdot$H$_2$O and Y$_2$(SO$_4$)$_3$$\cdot$8H$_2$O which means some of them exist diffusion phase at 873K such as Li$_2$SO$_4$ \cite{Li2SO4_1,Li2SO4_2}. 
The thermogravimetric analysis (TGA) shows that exhibit no significant weight loss up 1073K. 
It means that the melting point of LiY(SO$_4$)$_2$ is higher than 1073K while its synthesis temperature is below it. 
To our best knowledge, Li$_2$SO$_4$ was uncovered a first-order phase transition in Ref. \cite{Li2SO4_1,Li2SO4_2} and the rotation of SO$_4^{2-}$ was considered as a connection with ionic conductivity in Li$_2$SO$_4$ \cite{Li2SO4_rotation,Li2SO4_rotation1,Li2SO4_rotation2}. 
Considering the precursors are both sulfates it is a strong signal that it is synthesized by the superionic state. 
The authors also said that it exists losing the volatile Li slowly at 1000K \cite{LYSO} which is a common phenomenon in cathode synthesized so it is necessary to use extra 4-5\% Li source \cite{AM} to ensure stoichiometric ratio. 
In addition, as we discuss above, the Nonlinear Optics may have the connection with Superionic state.

\begin{figure}
\includegraphics[width=\textwidth]{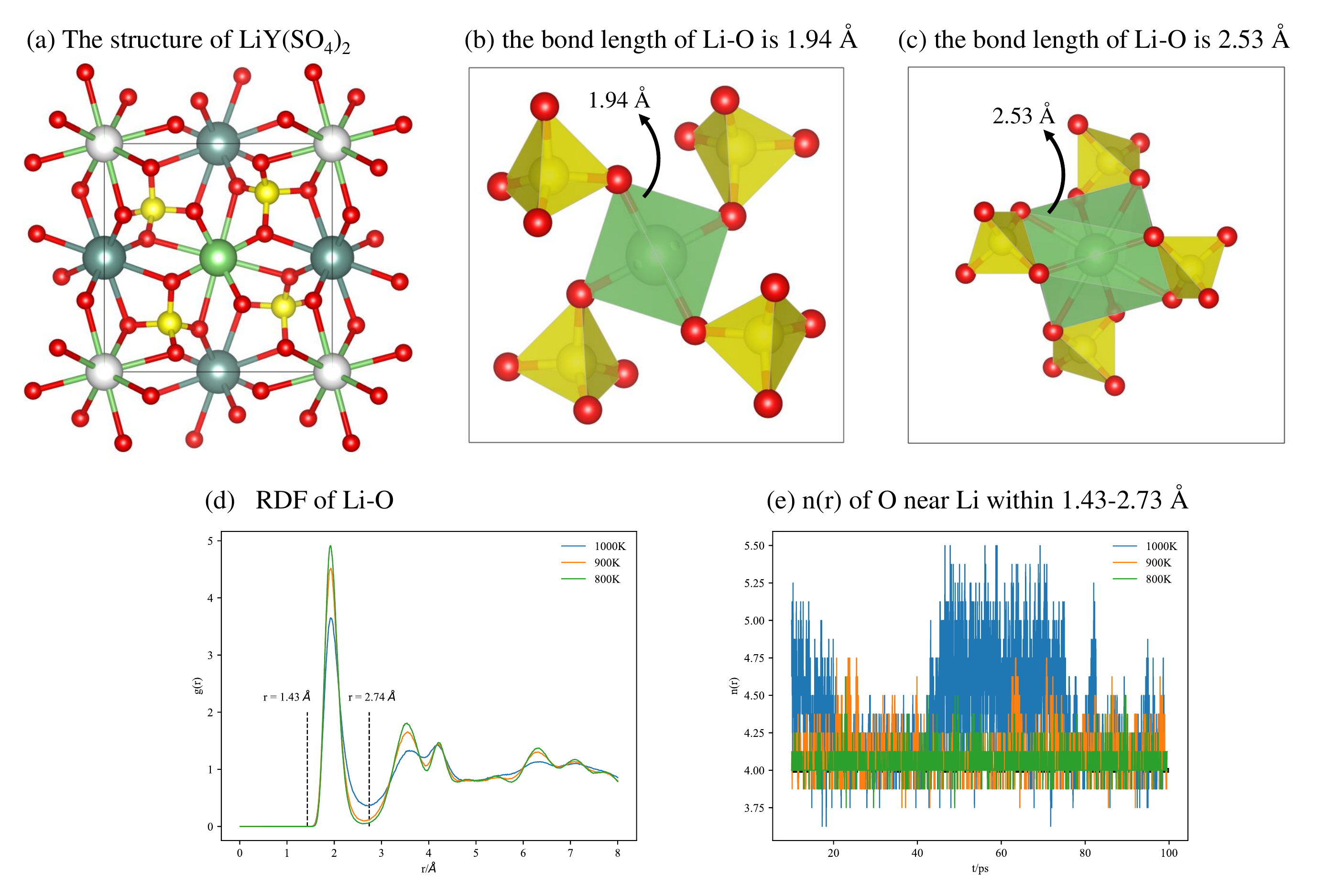}
\caption{\label{fig_str}(a) the structure of LiY(SO$_4$)$_2$ and the white atom is Li located at the interstitial sites. (b) the Li locates at ($\frac{1}{2}$,$\frac{1}{2}$,0) which is the lattice site. (c) the Li locates at (0,0,0) which is the interstitial site. (d) RDF of Li-O at 800K, 900K and 1000K. (e) n(r) of O near Li between 1.43 $\text{\AA}$ and 2.74 $\text{\AA}$ at 800K, 900K and 1000K.}
\end{figure}

\subsection{Structure Analyzing}
We first analyze the geometric structure of LiY(SO$_4$)$_2$. As shown in Figure \ref{fig_str}(a), the Li locates at the 2\textit{i} site ($\frac{1}{2}$,$\frac{1}{2}$,0) with four oxygen atoms surrounding as the nearest neighbor shown in Figure \ref{fig_str}(b) and the bond length of Li-O is 1.94 $\text{\AA}$.
There also exists a similar 2\textit{i} site (0,0,0) represented as white in Figure \ref{fig_str}(a) with eight oxygen atoms surrounding as the nearest neighbor shown in Figure \ref{fig_str}(c) and the bond length of Li-O is 2.53 $\text{\AA}$.
It means that if Li$^+$ diffuses from site ($\frac{1}{2}$,$\frac{1}{2}$,0) to the site (0,0,0), the Li-O the nearest neighbor numbers will increase double which maybe unstable because it needs more energy. 
The following analyze will reveal that the reorientation of SO$_4^{2-}$ will accompany with the diffusion of Li$^+$.

\subsection{\textit{Ab Initio} Calculation}
\begin{figure}
\includegraphics[width=\textwidth]{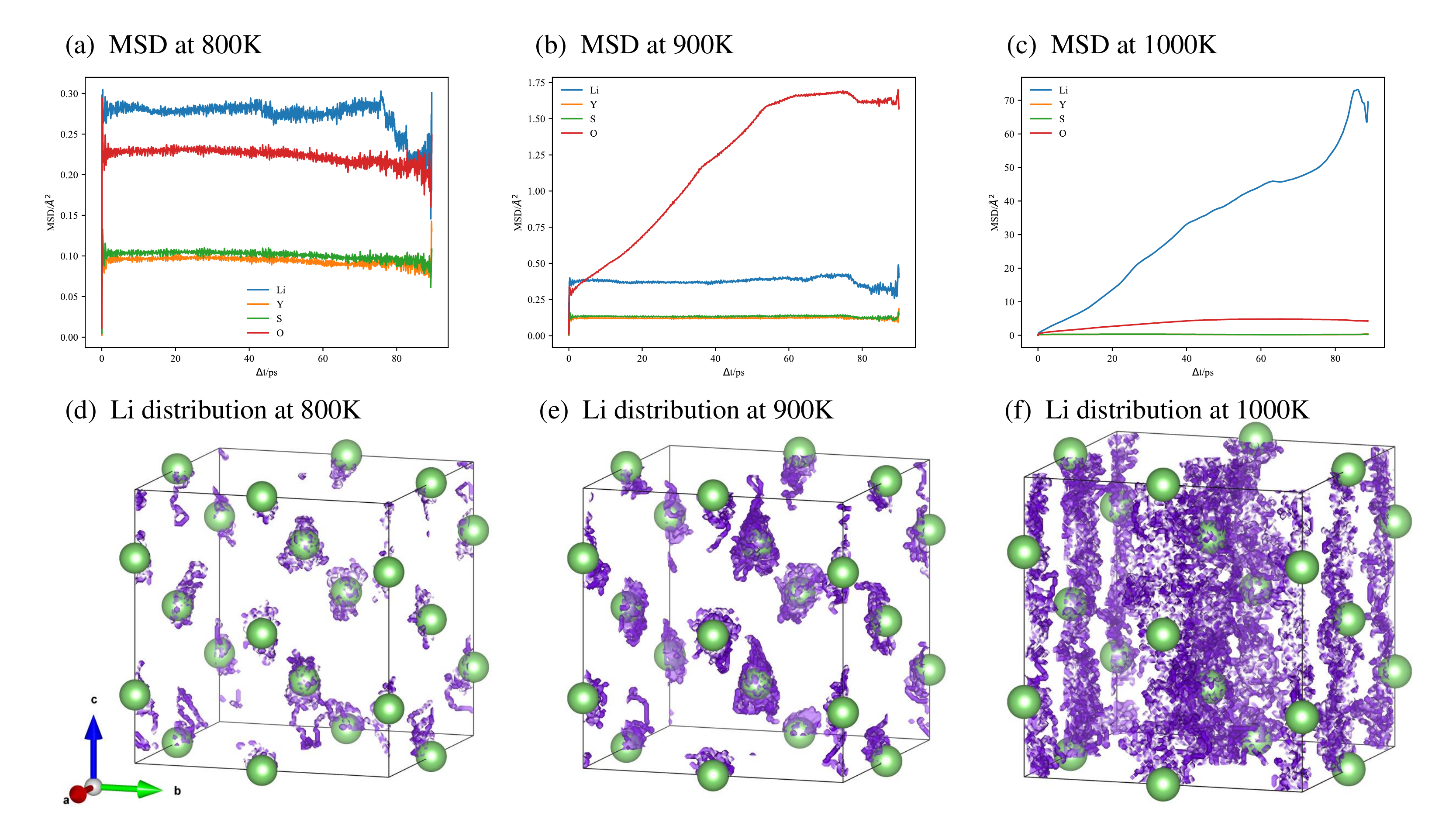}
\caption{\label{fig_msd}MSD for Li, Y, S, O in LiY(SO$_4$)$_2$ at (a) 800K, (b) 900K and (c) 1000K respectively; the Li distribution at (d) 800K, (e) 900K and (f) 1000K respectively.}
\end{figure}

We use \textit{ab initio} Molecular Dynamics (AIMD) with Nos\'e-Hoover thermostat \cite{Nose_1,Nose_2,Nose_3} by Vienna Ab initio Simulation Package (\textsc{\textbf{VASP}}) \cite{VASP_1} to simulation the NVT ensemble for a $\sqrt{2} \times \sqrt{2} \times 2$ LiY(SO$_4$)$_2$ supercell. 
We choose 800K, 900K and 1000K with the timestep of 1fs for 100ps to compare and the first 10ps was considered as the equilibrium process. 
It is clearly that at 800K all atoms have little displacements, at 900K few of O exist little displacements while at 1000K the Li atoms exist evident diffusion process as shown in Figure \ref{fig_msd}(a), (b) and (c) respectively from mean square displacement (MSD) and in Figure \ref{fig_msd}(d), (e) and (f) respectively from Li distribution. 
It indicates that at 800K the LiY(SO$_4$)$_2$ still maintains solid state and at 900K the O atoms exist few perturbation (we will discuss later) while 1000K the superionic state happens which means the LiY(SO$_4$)$_2$ exists superionic state synthesized at 873K. 
Considering Li$_2$SO$_4$ undergoes a first-order phase transition at 575 $^{\circ}$C and melts at 860 $^{\circ}$C \cite{Li2SO4_1,Li2SO4_2}, it is concluded that LiY(SO$_4$)$_2$ is synthesized through superionic states.\par

\begin{table}
\centering
\caption{\label{n_r} The proportion of nearest neighbor numbers within 1.43-2.74 $\text{\AA}$ for Li-O at 800K, 900K and 1000K}
\begin{tabular}{lllllll}
\hline
Temperature/K    & LiO$_3$ & LiO$_4$ & LiO$_5$ & LiO$_6$ & LiO$_7$ & LiO$_8$  \\
\hline
800  & 0.36\% & 96.57\% & 2.94\% & 0.12\% & 0.01\% & 0.00\%  \\
900  & 0.90\% & 91.78\% & 7.00\% & 0.32\% & 0.01\% & 0.00\%  \\
1000 & 3.39\% & 66.72\% & 24.72\% & 4.77\% & 0.38\% & 0.02\% \\
\hline
\end{tabular}
\end{table}

\subsection{Synthesized by Superionic State}
\begin{figure}[H]
    \includegraphics[width=\textwidth]{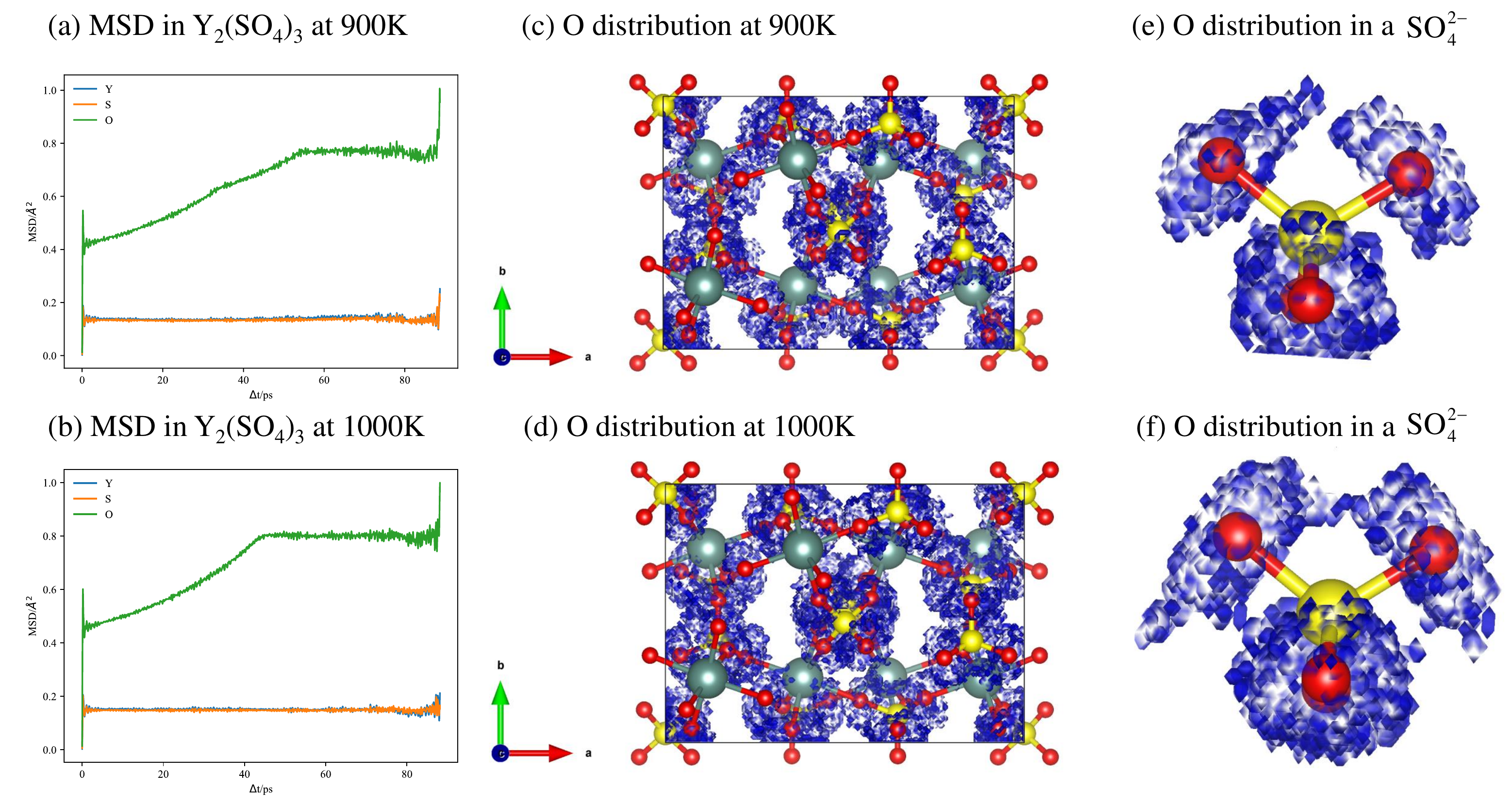}
    \caption{\label{fig_YSO}The MSD at (a) 900K and (b) 1000K and O distribution at (c) 900K and (d) 1000K and local O distribution at (e) 900K and (f) 1000K.}
\end{figure}

In order to explore the synthesis mechanism, we calculated the dynamics in Y$_2$(SO$_4$)$_3$ at 900K and 1000K.
It is evident that no O diffusion at 900K and 1000K in Y$_2$(SO$_4$)$_3$ from Figure \ref{fig_YSO}(a) and (b). 
However, the O rotation will be stronger in 1000K than 900K as shown in Figure \ref{fig_YSO}(e) and (f).
Considering the authors mixed them using an agate mortar and pestle thoroughly \cite{LYSO}, only a little ionic diffusion can induce the synthesis of LiY(SO$_4$)$_2$ because of the diffusion of Li$^+$ existing in Li$_2$SO$_4$ and LiY(SO$_4$)$_2$ and the rotation of O$^{2-}$ existing in Li$_2$SO$_4$ and Y$_2$(SO$_4$)$_3$. 
Milling accelerates the grain contact and reduces the grain size which increases the mixing of SO$_4^{2-}$ units from Li$_2$SO$_4$ and Y$_2$(SO$_4$)$_3$.
In fact, some superionic conductors such as Ag$_2$CdI$_4$ \cite{Ag2CdI4}, LiNbOCl$_4$ \cite{LNOC}, Pb$_{1-x}$Sn$_x$F$_2$ \cite{PbSnF2} and cathode materials such as Na$_{2+2x}$Fe$_{2-x}$(SO$_4$)$_3$ \cite{NaFeSO} and LiFePO$_4$F \cite{LFPF} are synthesized by milling and low temperature.

\subsection{Diffusion Mechanism in LiY(SO$_4$)$_2$}

In order to explore the mechanism of the diffusion of Li$^+$, we calculated the radial distribution functions (RDF) during 11-100ps at 800K, 900K and 1000K. 
As shown in Figure \ref{fig_str}(d), the atomics pairs of Li-O still maintain their isolated peaks at 800K and 900K while they become broadened because of the increasing of thermal fluctuations from 800K to 1000K which means it may own the character of liquid or glass. 
We select the nearest neighbor peaks as 1.43-2.74 $\text{\AA}$ for Li-O for further analyzing because the distances between two probable Li sites with the nearest neighbor O are 1.94 and 2.53 $\text{\AA}$ as shown in Figure \ref{fig_str}(b) and (c). 
We found at 800K and 900K the nearest neighbor numbers of O can almost maintain 4 while at 1000K they have a fluctuation between 4 and 5 in Figure \ref{fig_str}(e). 
The number of nearest neighbor O is mainly 4 as LiO$_4$ at 800K and 900K while it is 4 and 5 as LiO$_4$ and LiO$_5$ at 1000K as shown in Table \ref{n_r}. 
We notice the n(r) of O is not 8 which means the reorientation of SO$_4^{2-}$ or the fluctuations of O will accompany with the diffusion of Li$^+$ as same as in Introduction and our previous research \cite{wsy}.\par

\begin{figure}[H]
\includegraphics[width=\textwidth]{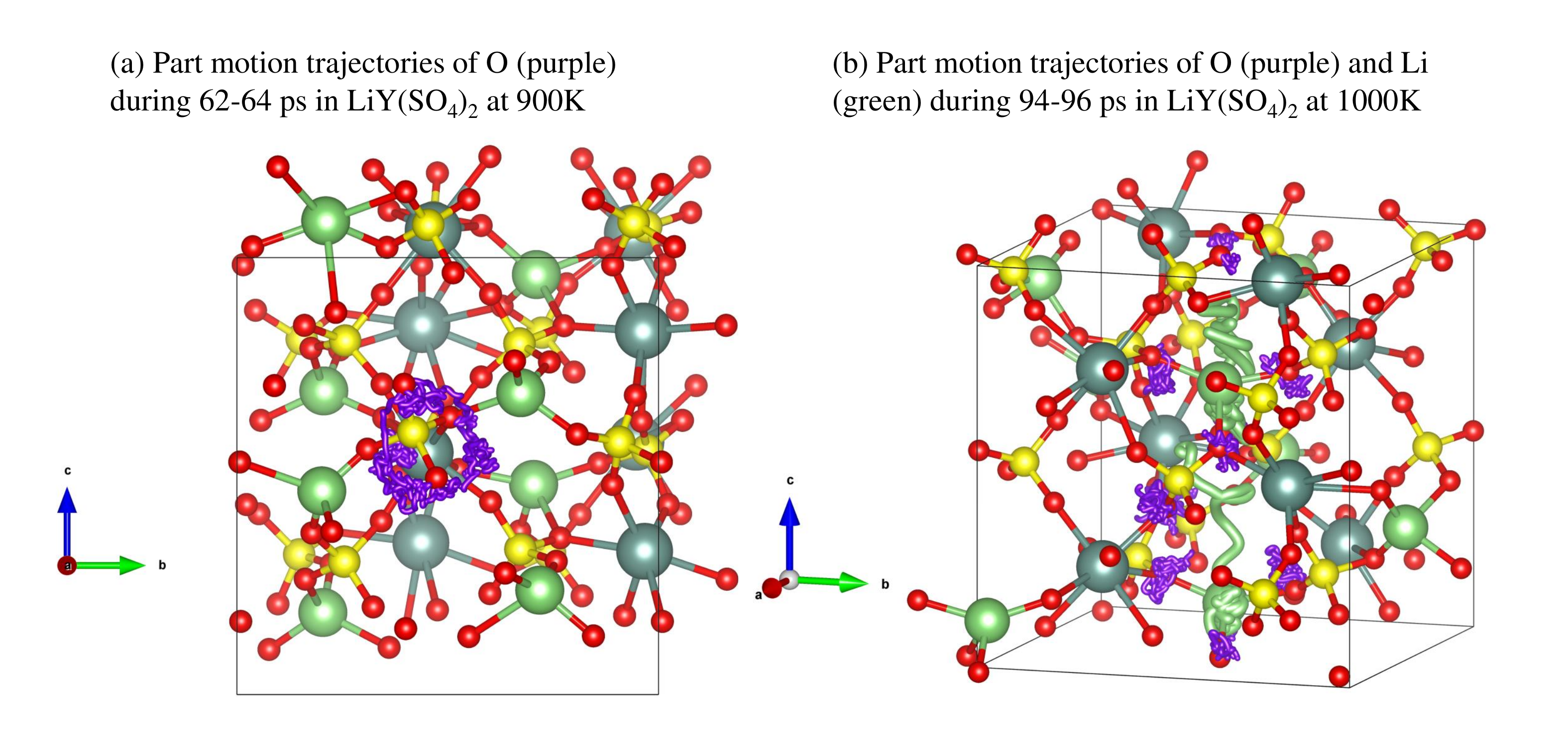}
\caption{\label{fig_trajectory}The motion trajectories of (a) O during 62-64 ps at 900K and (b) Li and O during 94-96 ps at 1000K in LiY(SO$_4$)$_2$, the trajectory of O and Li are labeled as purple and green respectively.}
\end{figure}

Some authors \cite{NYC,2022NRM,rotation_QENS,rotation_NMR,rotation_MD} think the rotation/reorientation of anion will be a method to enhance the cation ionic conductivity as shown in Na$_{3-x}$Y$_x$Zr$_{1-x}$Cl$_6$ \cite{NYC}. 
The phase is also called rotor phase. 
Although there exist abundant evidences by quasi-elastic neutron scattering \cite{rotation_QENS}, nuclear magnetic resonance \cite{rotation_NMR} and AIMD \cite{rotation_MD}, it seems there exists a question: what can be considered as reorientation or rotation? 
As shown in Figure \ref{fig_trajectory} we draw the trajectories of part atoms in LiY(SO$_4$)$_2$ at 900K/1000K. 
It is obviously that the MSD of O at 900K as shown in Figure \ref{fig_msd}(b) is caused by the $C_3$ rotation of SO$_4^{2-}$ from the results in Figure \ref{fig_trajectory}(a) while there is no signal for diffusion of Li$^+$. 
However, it seems only O atoms exist little disordered as shown in Figure \ref{fig_trajectory}(b) at 1000K accompany with the diffusion of Li$^+$ which is similar to our previous research Li$_2$BF$_5$ \cite{wsy}. 
It is hard to distinguish whether it belongs to rotation or at least it is disordered or reorientation. 
We can also conclude that the disordered O reduces the number of LiO$_{6/7/8}$ when diffusion of Li$^+$ as shown in Figure \ref{fig_str}(e) and Table \ref{n_r}.\par

\par

From the view of energy, superionic state transformation also belongs to the phase transformation and needs energy. 
The change from LiO$_4$ to LiO$_5$ even LiO$_{6/7/8}$ means the bond making and breaking. 
If the number of bond change is large, the system is equivalent to battery which can store thermal energy. 
The reorientation of SO$_4^{2-}$ or the diffusion of Li$^+$ means the fluctuation of energy. 
More bonds change, more energy needs and higher transformation temperature is. 
In order to reduce energy or superionic state transformation temperature, the reorientation of SO$_4^{2-}$ will create more LiO$_4$ and LiO$_5$ when Li$^+$ diffuses to reduce energy. 
So, the reorientation of O is a method to reduce the higher energy configuration in LiO$_x$ rather than a signal of diffusion of Li$^+$.

\subsection{Discussion}
Our research presents threefold meanings. 
First, we conclude the character of superionic state in literature: 'liquid-like' or 'sublattice/partial melting'. 
Although there exist different domains pay attention to superionic state, we can converge it based on these keywords. 
Second, apart from the keywords like 'liquid-like' or 'sublattice/partial melting', the synthesis method and other properties can be another indicator for seeking new superionic materials from literatures. 
From the view of physics, the superionic state represents partial liquid in solid so the properties locating between solid state and liquid state must become the focus. 
We can use it to find new superionic materials as same as this research. Compared with high throughput calculations \cite{JPCL_2011,Curtarolo_CMS_2010} with known structures from the literatures, our research acquires not only the structure of LiY(SO$_4$)$_2$ but also the stability and wide bandgap from Ref. \cite{LYSO} which are two key indicators for solid state electrolytes and it saves large computing resource. 
We must emphasize that this method doesn’t need any extra experiments because the structure parameters and thermodynamics property are basic contents in synthesis literatures. 
Third, our research discuss the micromechanism of the superionic state in LiY(SO$_4$)$_2$: the reorientation of SO$_4^{2-}$ accompanied with the diffusion of Li$^+$ for reducing the bond making. 
It is beneficial to not only understanding the micromechnism of diffusion of Li$^+$ but also assisting synthesizing new materials at low temperature.

\section{\label{con}Conclusion}
In this Letter, we present an example of excavating new superionic materials from the literature. 
Our research presents a new method on the condensed matter physics and material science. 
It is not only in favor of classifying the similar research fields such as superionic states but also presents a new insight to discovery new materials in other fields. 
We also discuss the mechanism of superionic state in LiY(SO$_4$)$_2$ and find the reorientation of SO$_4^{2-}$ accompanied with the diffusion of Li$^+$.

\begin{acknowledgement}
We acknowledge the Tianjin Supercomputer Center for providing computing resources and this research was funded by the National Natural Science Foundation of China (grant number 52022106 and 52172258) and the Informatization Plan of Chinese Academy of Sciences (grant number CAS-WX2021SF-0102).
\end{acknowledgement}


\bibliography{LiYSO_4_2}
\end{document}